\documentclass[11pt,preprint]{aastex}

\usepackage{graphicx}

\shortauthors{Hughes \& Wallerstein}
\shorttitle{Photometric Metallicities}

\begin{document}

\title{Determining Photometric Metallicities of dSph Stellar Populations}

%
%

\author{Joanne Hughes}
\affil{Physics Department, Seattle University, Seattle, WA 98122}
\email{jhughes@seattleu.edu}
\and
\author{George Wallerstein}
\affil{Astronomy Department, University of Washington, Box 351580, Seattle,
WA 98195-1580}

%
%
\email{walleg@u.washington.edu}

\begin{abstract}
If there are so few upper red-giant branch stars in the SDSS-discovered dwarf galaxies, how can we find the true population structure without extensive spectroscopy? We review recent  photometric and spectroscopic studies of the Ultra-Faint Dwarf Galaxies, and determine a new method of estimating $[Fe/H]$ with a combination of Washington and Str\"{o}mgren filters, using Bo\"{o}tes I dSph  as an example. We can use the  $CT_1by$ filters alone to achieve $0.3$~dex resolution in 
[Fe/H], and 0.5~Gyr resolution in age. Both the Washington and  Str\"{o}mgren filters, $C$ and $v$, 
are sensitive to CN-variations; however, in stars with a large deficiency of heavy elements the CN 
bands are weak and not important.  The [Fe/H]-sensitivity of the
Washington and Str\"{o}mgren combination is at least twice as great as the SDSS filters, and 
this work maintains that resolution on the lower red-giant branch, where other calibrations fail.
\end{abstract}

\section{Introduction}

This article summarizes and explores  a recent project in which we found photometric metallicities within SDSS-discovered dSph galaxies, using novel filter and color combinations. For more details, see Hughes, Wallerstein \& Dotter (2011; hereafter, HWD) and  Hughes, Wallerstein \& Bossi (2008; hereafter, HWB). 
The dark matter dominated ($M/L>100$), ultra-faint dwarf galaxies (UFDs; Willman 2010 reviews the SDSS search methods), are  the  Òleast luminous galaxiesÓ, which can be as faint as $10^{-7}$ times the luminosity of the Milky Way (range: $300<L_\odot<100,000$).

We have studied the age and metallicity distributions in several dwarf spheroidal (dSph) systems, using Bo\"{o}tes I as the proving ground (HWB; HWD). Frebel, Simon \& Kirby (2011)  studied the chemical composition of several UFDs with high-resolution spectroscopy.  A recent paper by Lai et al. (2011) used low resolution spectra, SDSS and other available filters to determine [Fe/H], [C/Fe], and [$\alpha$/Fe] for each star, utilizing a 
new version of the SEGUE Stellar Parameter Pipeline (SSPP; Lee et al. 2008a,b), named the n-SSPP (the method for non-SEGUE data). 

In Boo I,  enough data exist on the stellar population to make a comparison among these different methods
of assessing the chemical composition.
Martin et al.'s (2007) CaT data  suggest the stars in our sample have a range in [Fe/H] of  over $> 1.5$ dex, which is found to be even greater from the n-SSPP method and high-resolution spectra (Lai et al. 2011; Norris et al. 2010b; Feltzing et al. 2009; HWB; Martin et al. 2007). 

Lai et al. (2011) find the 
greatest range in [Fe/H] at $>2.0-2.5$ dex, and a mean $[Fe/H]=-2.59$. HWB, using Washington photometry, find $[Fe/H]=-2.1$, and a range $> 1.0$ dex in the central field. Martin et al. (2007) found the same mean value as HWB with the calcium triplet (CaT) method (30 objects). It is known that 
the CaT calibration may skew to higher [Fe/H]-values at the lower-metallicity end, below $[Fe/H]\sim-2.0$ (Kirby et al. 2008). The Geisler \& Sarajedini (1999; hereafter, GS99) standard giant branches, in the Washington filters, are used to calibrate HWB's estimate.  Siegel (2006) notes that Boo I's stellar population is similar to that of M92
(which HWB and HWD also found), and we note that recently, M92 and M15 are regarded as the most metal poor 
globular clusters at $[Fe/H]\sim -2.3$. GS99 discuss the metallicity scales of  Zinn \& West (1985) and that of Carretta \& Gratton (1997), preferring the latter. GS99 also cite Rutledge, Hesser, \& Stetson's (1997) study of calcium-triplet strengths, in  support the of Caretta \& Gratton's (1997) scale.
However, GS99 then show that M15 returns $[Fe/H]=-2.15$ on the Zinn \& West (1985) scale, but
$-$2.02 on that of Carretta \& Gratton (1997). Within the uncertainties, this alone explains the difference
in mean [Fe/H] between the Washington photometry and the SDSS data. The Washington filters
and the GS99 standard giant branches are meant to return the CaT-matched metallicity scale of Carretta \& Gratton (1997).

\section{Observations}

We observed several SDSS-discovered dSphs in 2007-2011, with the Apache Point Observatory (APO) 3.5-m telescope's SPIcam Imager (FOV: $4.8^\prime \times 4.8^\prime$), using a combination of Washington and Str\"{o}mgren colors/indices (Str\"{o}mgren 1956;
Crawford \& Mander 1966). HWB's data shows that Bo\"{o}tes I has a spread in  [Fe/H] of  $> 1.0$ dex, using the red giant branch (RGB) and the main sequence turn-off (MSTO). For this project (detailed in HWD), we obtained photometry of a field central to the Boo I (dSph) galaxy, which was first discovered as a stellar over-density in the Sloan Digital Sky Survey (Belokurov et al. 2006). We used $CT_1T_2vby$ filters, where the $CT_1T_2$ data was published in HWB. HWD also compares the Str\"{o}mgren and SDSS bands. We reduced the images and performed photometry on the objects in each filter using the software in IRAF (and its version of DAOPHOT).

\section{Practical Filter Sets for dSphs}

With an average $[Fe/H] \sim -2.5$, some stars in UMa II, Segue 1, Boo I are even below $[Fe/H]=-3.5$
(Norris et al. 2010a; Norris et al. 2010b; Frebel et al. 2010). Recent papers have explored the best color-pairs to use for age and metallicity studies 
(e.g., Li \& Han 2008; Holtzman et al. 2011). However, much of the work is theoretical
and involves testing on local, highly populated globular clusters.

Figure~1 shows the transmission curves for the filters given in Table~1 (also see: Bessell 2005), from the CTIO website\footnote{http://www.ctio.noao.edu/instruments/filters/index.html}, with the ATLAS9\footnote{http://wwwuser.oat.ts.astro.it/castelli/grids.html} model flux density for a star with $T_{eff}=4750 K$, $[Fe/H]=-2.5$, $[\alpha/Fe]=+0.4$, $\log g=1.5$.  The best photometric system designed for separating stars by metallicity is considered to be the intermediate-band Str\"{o}mgren photometry. The distant RGB stars in the dSphs are very faint at Str\"{o}mgren-u, and only the 8 brightest members are detected at SDSS-u. Thus, we are unable to obtain the surface gravity-sensitive $c_1$-index, where $c_1=(u-v)-(v-b)$. The metallicity of the stars is sensitive to the $m_1$-index, where $m_1=(v-b)-(b-y)$. The color (b-y) is a measure of the temperature and (v-b) is a measure of metallic line blanketing (see Figures 1 and 2). Several groups have mapped the Str\"{o}mgren metallicity index to [Fe/H] (e.g., Hilker 2000; Calamida et al. 2007; 2009) and find that calibrations fail for the RGB stars at $(b - y) < 0.5$ for all schemes. Faria et al. (2007) made the comment that metal-rich and metal-poor stars are mixed
together on the lower-RGB, which they say is likely due to the larger photometric errors. Although 
this is partially accurate, the Figure~1 and Figure~2 provide the real answer. The $m_1$-index loses sensitivity as the line absorption in $v$ becomes equal to the line absorption in $b$. In other words, the difference in line absorption between $b$ and $v$ becomes equal to the difference in line absorption between $b$ and $y$. As stars get fainter on the RGB, the surface temperature rises and the lines get weaker (also see: Onehag et al. 2009; Arnadottir, Feltzing \& Lundstrom 2010, and references therein).

\begin{figure}
\centering
\includegraphics[scale=0.5]{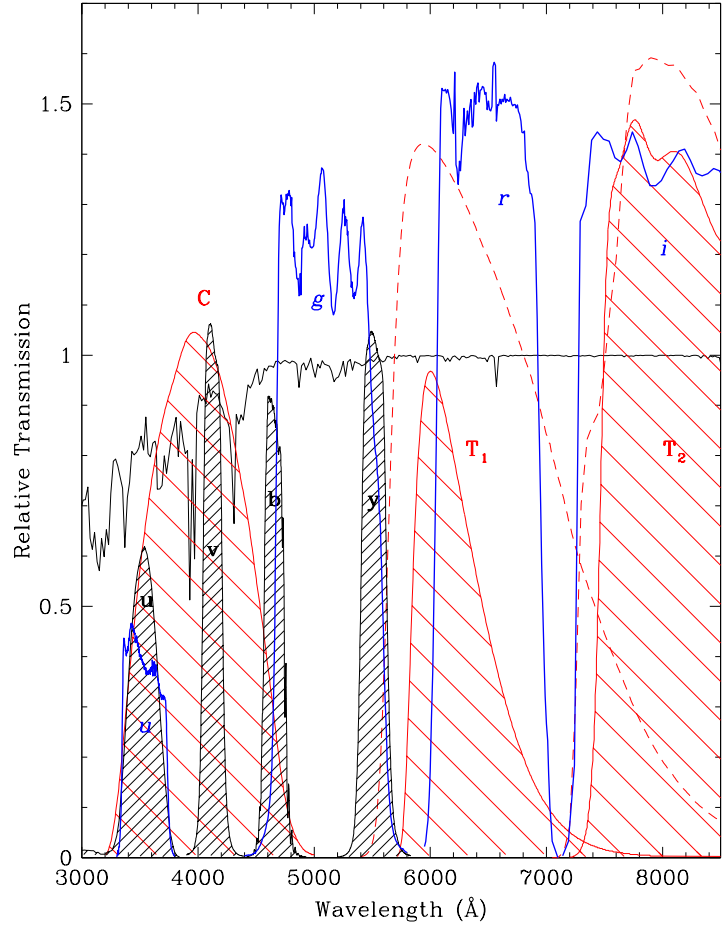}
\vskip0pt
\caption{Transmission curves for the filters given in Table~1, from the CTIO website. We also show
the ATLAS9 model flux density for $T_{eff}=4750 K$, $[Fe/H]=-2.5$, $[\alpha/Fe]=+0.4$, $\log g=1.5$. Str\"{o}mgren filters (including u) are shown as shaded black curves. Washington filters are shown in shaded red, with the R and I filters as dashed lines. SDSS filters are shown in blue.}
\label{fig1}
\end{figure}

\begin{deluxetable}{lccc}
\tabletypesize{\normalsize}
\tablecaption{Filters Used for our dSph Studies}
\tablewidth{0pt}
\tablehead{\colhead{Filter}   & \colhead{Central $\lambda$(\AA )} &
       \colhead{Width (\AA )} & \colhead{Remarks} } 
\startdata
 v &  4100 & 190& Str\"{o}mgren \\
b &  4690 & 180& Str\"{o}mgren\\
 y &  5480 & 230& Str\"{o}mgren \\
 C & 3980& 1100& Washington \\
$T_1$ &  6390& 800& Use R\\
 $T_2$ & 8050& 1500& Use I\\
 u &  3550& 570& SDSS \\
 g & 4690& 1390& SDSS \\
 r & 6160& 1370& SDSS \\
 i &  7480& 153& SDSS \\
\enddata
\end{deluxetable}

\begin{figure}
\centering
\includegraphics[scale=0.5]{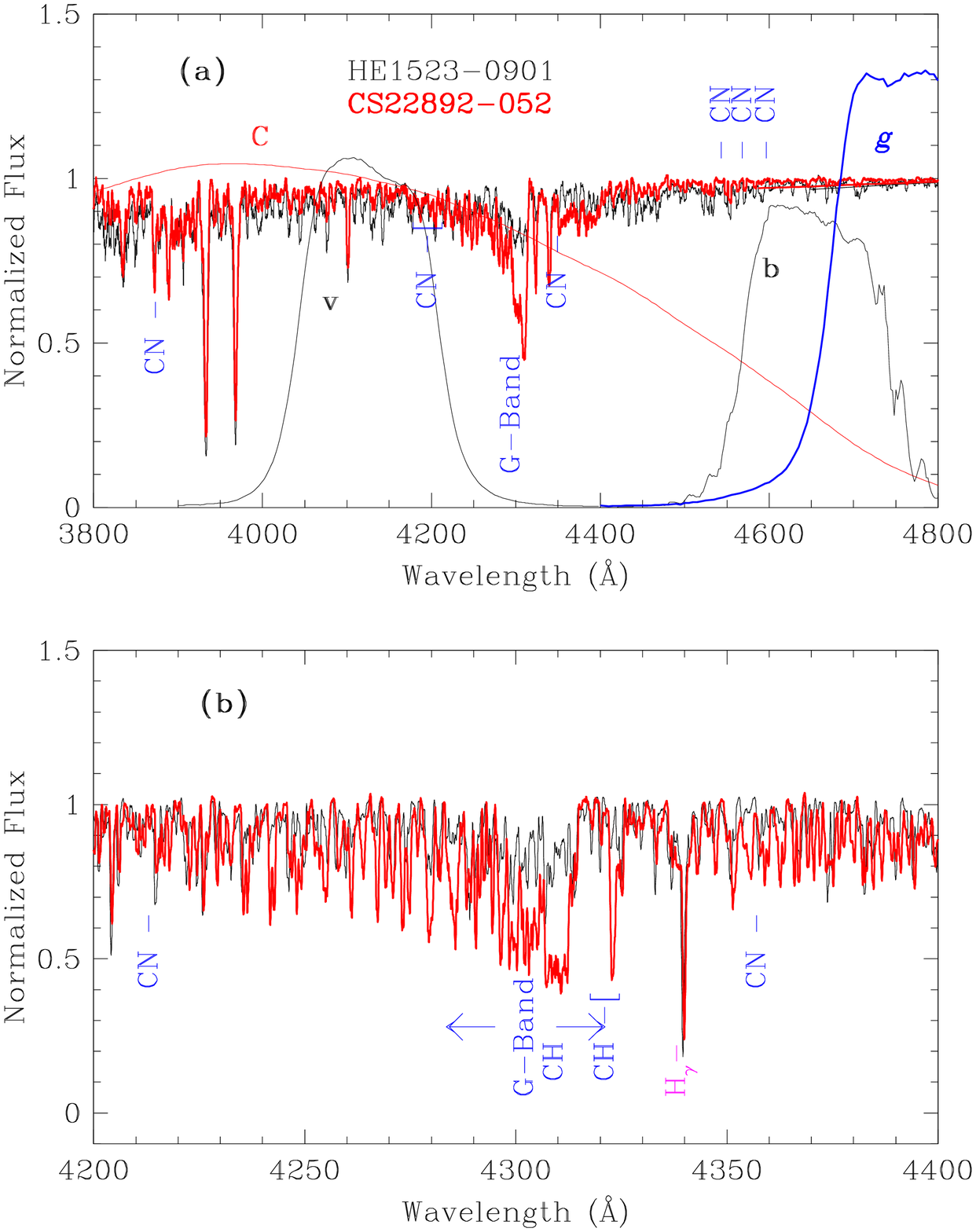}
\vskip0pt
\caption{\bf (a) \rm Normalized flux plots for HE1523-0901 (black) and CS22892-052 (red) (from 3800-4800\AA )with filter
transmission curves for $C$ (red), $v$ (black), $b$ (black), and $g$ (blue) filters. We also note the major CN and CH features. The original resolution has been smoothed to show the $C$-sensitive absorption.
\bf (b) \rm The normalized flux curves for HE1523-0901 (black) and CS22892-052 (red) (from 4200-4400\AA ), with major CN \& CH spectral features marked. }
\label{fig2}
\end{figure}

Also from Figure~1, we can see  the advantages that the Washington filters provide (GS99). The $C$-filter
covers the metallicity-defining lines contained in the narrower $v$-filter and part of the $b$-filter. The $C$-band also includes the 
surface-gravity sensitive Str\"{o}mgren-$u$ and SDSS-$u$. Thus, the color $(C-T_1)$ should be able to give information on $T_{eff}$, $[Fe/H]$, $[\alpha/Fe]$, and $\log g$. The Str\"{o}mgren filters are only better than Washington bands if you have a well-populated upper RGB stars, or  the system is close enough to have good photometry below the SGB, where the isochrones separate, in the Str\"{o}mgren system. The more-commonly used broadband $R-$ and $I$-filters can be converted linearly to Washington $T_1$ and $T_2$, but with less observing time needed. The $C$-filter is broader than the $B$-band, and is more sensitive to line-blanketing, and it is a much better filter choice than $B$ or  Str\"{o}mgren-$v$ for determining metallicity in faint, distant galaxies.

As mentioned by Sneden et al. (2003), metallicity is usually synonymous with [Fe/H], but
other elements may be inhomogeneously-variable in dSphs as well as the Milky Way's halo.
In Figure~2, we use 2 metal-poor stars to illustrate the sensitivity of the filters in Figure~1 to C-enhancement. HE 1523Ð0901 (black: Frebel et al. 2007) is an r-process-enhanced metal-poor star 
with $[Fe/H]\approx -3.0$, $[C/Fe]=-0.3$, $\log T_{eff}=4650K$, and $log \;g=1.0$. CS 22892-052 (red) is also an r-process rich object (Sneden et al. 2003; Sneden et al. 2009; Cowan et al. 2011) with $[Fe/H]\approx -3.0$, $[C/Fe]\approx 1.0$, $\log T_{eff}=4800K$, $log \;g=1.5$, and
$[\alpha /Fe]\approx +0.3$. The change in the CH-caused G-band is apparent, and CN/CH features
affect the $C$, $v$, and $b$ filters, but the SDSS $g$-band is relatively clear of contamination, but is also not generally very sensitive to metallicity. The spectra shown in Figure~2 were provided kindly by Anna Frebel (private communication).

In Figure~3, we show we show color-magnitude diagrams (CMDs) used for calibration of Boo I to M92 (cyan points: private communication, F. Grundahl; see Grundahl et al.  2000).
The dark blue line is the Dartmouth isochrone (Dotter et al. 2008) which fits well with a recent study by di Cecco et al. (2010), $DM=14.74$, $[Fe/H]= -2.32$, $[\alpha /Fe]=0.3$ and $Y =0.248$, and $age=11\pm1.5$Gyr. M92 has $E(B-V)=0.025$ and $DM=14.74$, and Boo I is taken to have $E(B-V)=0.02$ and $DM=19.11$.

\begin{figure}
\centering
\includegraphics[scale=0.52]{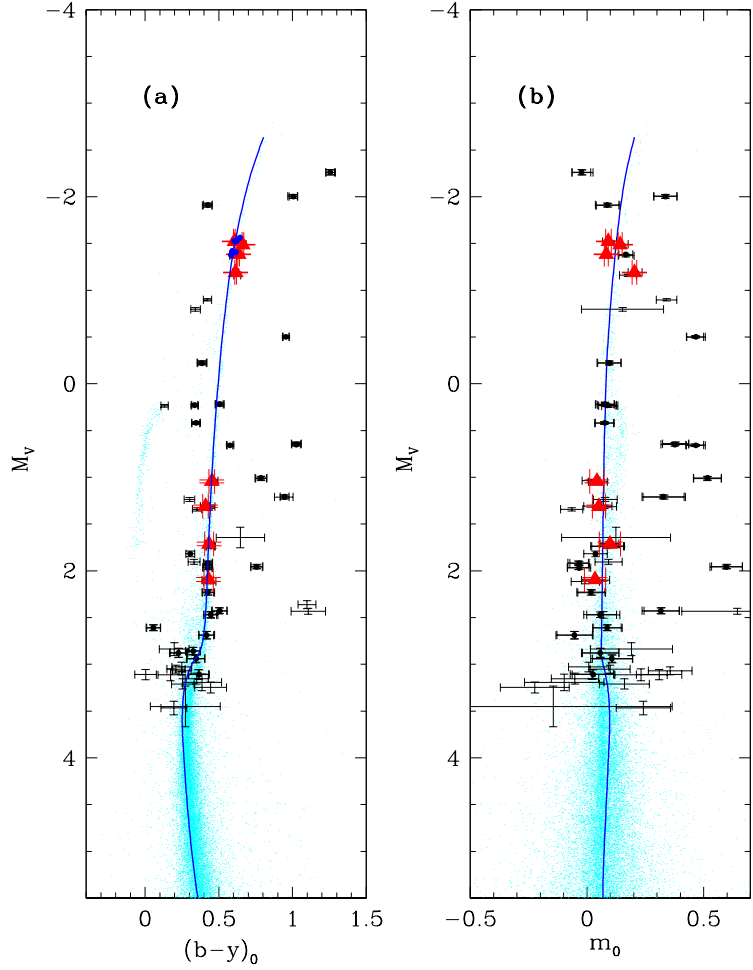}
\vskip0pt
\caption{\bf (a) \rm $M_V \; vs. \; (b-y)$ Str\"{o}mgren CMD for M92 (cyan points), Boo I Str\"{o}mgren only (black points),
Str\"{o}mgren and Washington objects (black filled circles), and proper-motion members with
Str\"{o}mgren, Washington and SDSS magnitudes (red filled triangles). 
The dark blue line is the Dartmouth isochrone corresponding to $[Fe/H]=-2.25$, $[\alpha /Fe]=0.3$ and an age of 11~Gyr. M92 has $E(B-V)=0.025$ and $DM=14.74$, and Boo I has $E(B-V)=0.02$ and $DM=19.11$.
(b) \rm  $M_V \; vs. \; m_0$ Str\"{o}mgren CMD for M92 (cyan points), Boo I Str\"{o}mgren only (black points),
Str\"{o}mgren and Washington objects (black filled circles), and proper-motion members with
Str\"{o}mgren, Washington and SDSS magnitudes (red filled triangles). The dark blue line is the Dartmouth isochrone corresponding to $[Fe/H]=-2.25$, $[\alpha /Fe]=0.3$, and an age of 11~Gyr.}
\label{fig3}
\end{figure}

\begin{figure}
\centering
\includegraphics[scale=0.6]{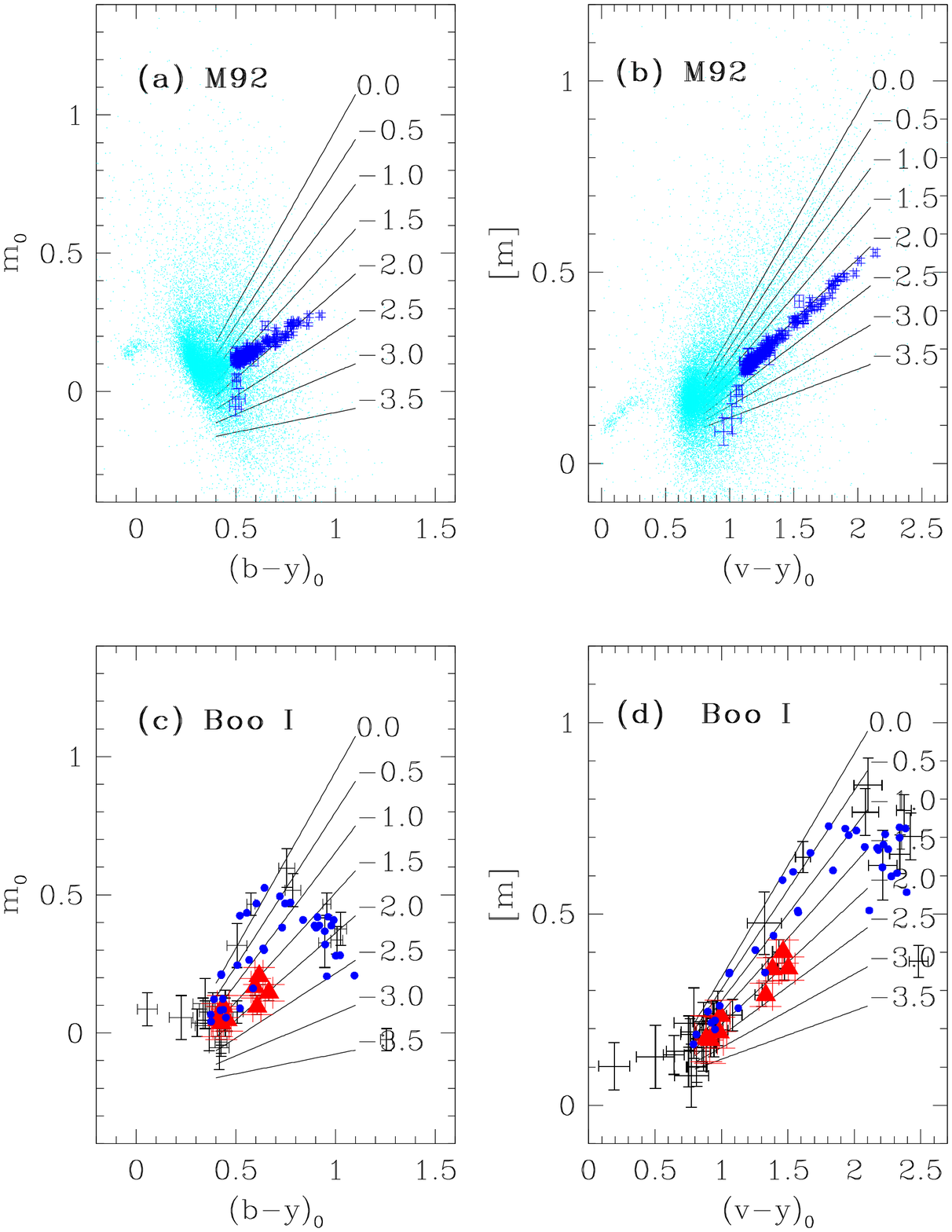}
\vskip0pt
\caption{\bf (a) \rm $m_0=(v-b)_0-(b-y)_0$ for M92 RGB stars (blue points, F.Grundahl, private communication), and the rest of the cluster stars (cyan). The calibration lines of constant
[Fe/H] come from Hilker (2000). \bf (b) \rm For the same M92 sample, we show
 $[m]=m_1+0.3(b-y)$, the reddening-free index, plotted with $(v-y)_0$. Calibration from Calamida et al. (2007). 
 \bf (c)  \rm $m_0=(v-b)_0-(b-y)_0$ for the HWD sample, with the Hilker (2000) calibration. In total, 59 objects were detected
 in $vby$ filters, shown as black points. The TRILEGAL code was used to generate a sample
 of foreground stars, shown as blue filled-circles. The 8 RGB stars from HWD are shown as red triangles. 
\bf (d) \rm  $[m]=m_1+0.3(b-y)\; vs. \; (v-y)_0$ for the sample of Boo I stars, with the Calamida et al. (2007) calibration.
}
\label{fig4}
\end{figure}

In Figure~4a and 4b,  we show color-color plots and [Fe/H]-calibrations for M92 (cyan points). The blue points  are the M92 RGB stars above the horizontal branch (HB). Having the same type of cool, metal-poor RGB as the dSph 
population, these plots illustrate the loss of metallicity resolution on the lower-RGB in the
Str\"{o}mgren system. HWB used a statistical cleaning method to remove foreground stars which contaminated the Boo I population. We used the TRILEGAL code\footnote{http://stev.oapd.inaf.it/cgi-bin/trilegal}  to generate a field of artificial stars at the correct galactic latitude, for the same magnitude limits as our dSph field. For each star in our field, we generate a probability that it is a dSph member, based on its colors and the number of neighboring field stars in a CMD. Figure~4c and 4d show our Boo I data (black points with error bars) and the TRILEGAL-generated artificial stars (blue circles). The red triangles are the bright RGB stars with SDSS-colors. The Str\"{o}mgren filters are well-suited to separate the dSph population from the
foreground stars.

\section{Discussion}

From Figure~5, we see that $(C -T_1)$ widens the separation of the giant branches of different metallicities, giving a resolution for RGB fiducials of $\sim 0.15$ dex, while the reddening sensitivity of the Washington filters is half that of the $(V - I)$ color. One of us (GW) defined the Washington system, which was developed by Canterna (1976), and Geisler (1996) defined CCD standard fields for the system. However, the Str\"{o}mgren system is more sensitive to low metallicities, $[Fe/H] < -2.0$, but requires longer integration times. In
HWB, we found that Washington filters spread out the stars at the MSTO, and we have found that $(C - T_1)$ is more effective than the SDSS-colors $(g - i)$ or $(g - r)$. We can also see that the best SDSS color (Figure~5b) for metallicity resolution is $(g-r)$, but that the SDSS photometry is not sensitive enough to this difference in colors to distinguish Boo~I's level of metallicity spread with the photometric errors of the SDSS.
 
\begin{figure}
\centering
\includegraphics[scale=0.6]{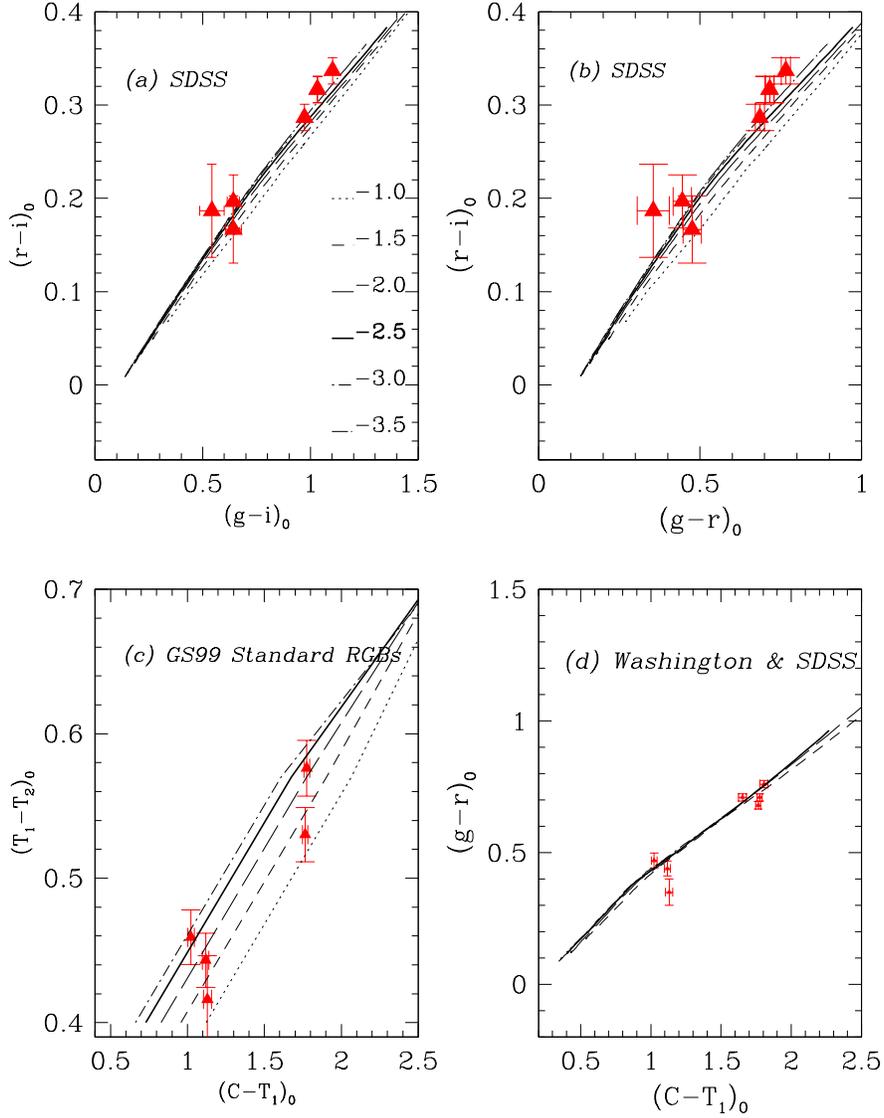}
\vskip0pt
\caption{Color-color plots using a mixture of SDSS and Washington filters to show the [Fe/H]-sensitivity (HWD). In all diagrams, the red triangles are the 8 RGB stars from HWD.  \bf (a) \rm $(r-i)_0 \; vs.\;  (g-i)_0$ with Dartmouth models.  \bf (b) \rm $(r-i)_0 \; vs.\;  (g-r)_0$ with Dartmouth models.
 \bf (c) \rm Washington colors with GS99 standard giant branches (HWB).
\bf (d) \rm  Combining metallicity sensitive colors $(g-r)$ and $(C-T_1)$ just results in a temperature and surface gravity index.}
\label{fig5}
\end{figure}

In Figure~6, we compare the Str\"{o}mgren and Washington filters, and construct two new indices: 
$m_*=(C-T_1)_0-(T_1-T_2)_0$ and $m_{**}=(C-b)_0-(b-y)_0$. The motivation is to avoid the 
collapse of the metallicity sensitivity of the $m_1$-index on the lower-RGB, and to attempt 
to replace the $v$-filter with the broader $C$-filter. We see that the most successful combination,
which maintains reasonable [Fe/H]-resolution over the whole RGB whilst allowing for increasing
photometric uncertainties towards the lower-RGB, is shown in Figure~6h, with $m_{**} \; vs.\;  (C-T_1)_0$. This result allows us to use 4 filters, $CT_1by$, which saves observing time and 
keeps $\sim  0.3$~dex [Fe/H]-resolution for stars with $-1.5<[Fe/H]<-4.0$. Figure~6i shows that 
we could use $Cby$ for metallicity estimates  $-1.5<[Fe/H]<-4.0$, but these color-color plots are not sensitive to age on the RGB. Figs.6a, b \& c show that $m_0$ will remain preferred for systems with $-1.0<[Fe/H]<-2.0$.

\begin{figure}
\centering
\includegraphics[scale=0.6]{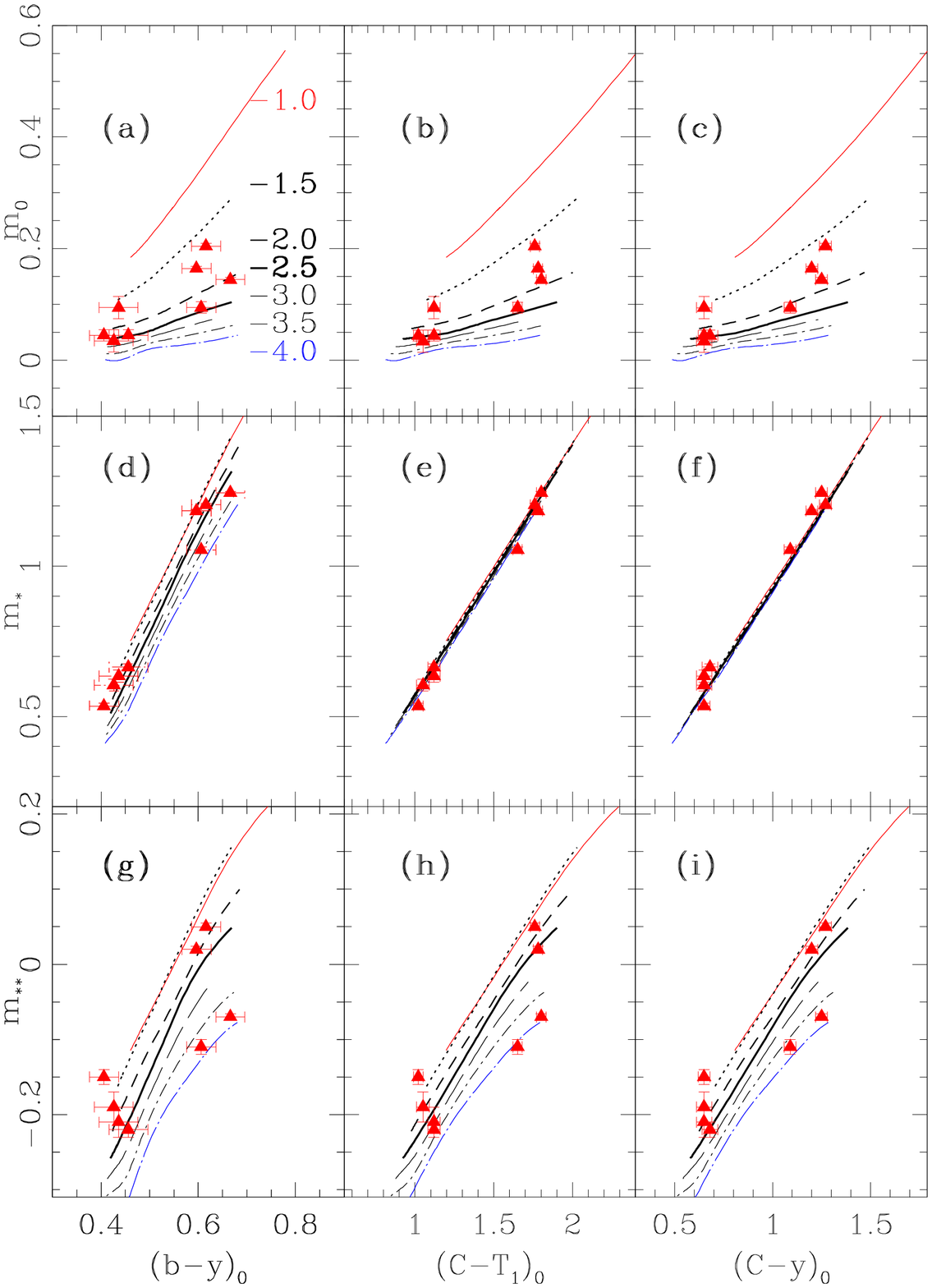}
\vskip0pt
\caption{Color-color plots using a mixture of Str\"{o}mgren and Washington filters to show the [Fe/H]-sensitivity (HWD). In all plots, the red triangles are the 8 RGB stars from HWD, and we show the Dartmouth models from $-1.0>[Fe/H]>-4.0$. We define: $m_*=(C-T_1)_0-(T_1-T_2)_0$ and $m_{**}=(C-b)_0-(b-y)_0$, with dereddened $m_1$ as $m_0=(v-b)_0-(b-y)_0$.
 \bf (a) \rm $m_0 \; vs.\;  (b-y)_0$.
 \bf (b) \rm $m_0 \; vs.\;  (C-T_1)_0$.
 \bf (c) \rm $m_0 \; vs.\;  (C-y)_0$.
\bf (d) \rm $m_* \; vs.\;  (b-y)_0$.
 \bf (e) \rm $m_* \; vs.\;  (C-T_1)_0$.
 \bf (f) \rm $m_* \; vs.\;  (C-y)_0$.
  \bf (g) \rm $m_{**} \; vs.\;  (b-y)_0$.
 \bf (h) \rm $m_{**} \; vs.\;  (C-T_1)_0$.
 \bf (i) \rm $m_{**}\; vs.\;  (C-y)_0$.
}
\label{fig6}
\end{figure}

In Figure~8a \& b, we compare the Str\"{o}mgren and Washington systems, respectively. This work will be expanded in HWD, but a simple closed-box chemical evolution model is run for  a total population of a few thousand  stars with a conservative range of $-1.0<[Fe/H]<-3.5$, ages $10-12$~Gyr, where we added  2\% photometric
uncertainties on the RGB, and we expect it to rise to at least 5\% at the MSTO.
We need at least 2\% photometry at the MSTO to determine if there are age spreads present, since the isochrones  exhibit a change of $\sim 0.06$ mag in $(C - T_1)$ for each Gyr in age (better than $(B-I)$ also). We show that there would have to be much deeper Str\"{o}mgren photometry for
any age spread to be seen, but that the metallicity spread is maintained. In the Washington system,
it is clear that this level of photometric uncertainty would reveal an age spread of $> 1$~Gyr, and no age-spread is observed.

\begin{figure}
\centering
\includegraphics[scale=0.6]{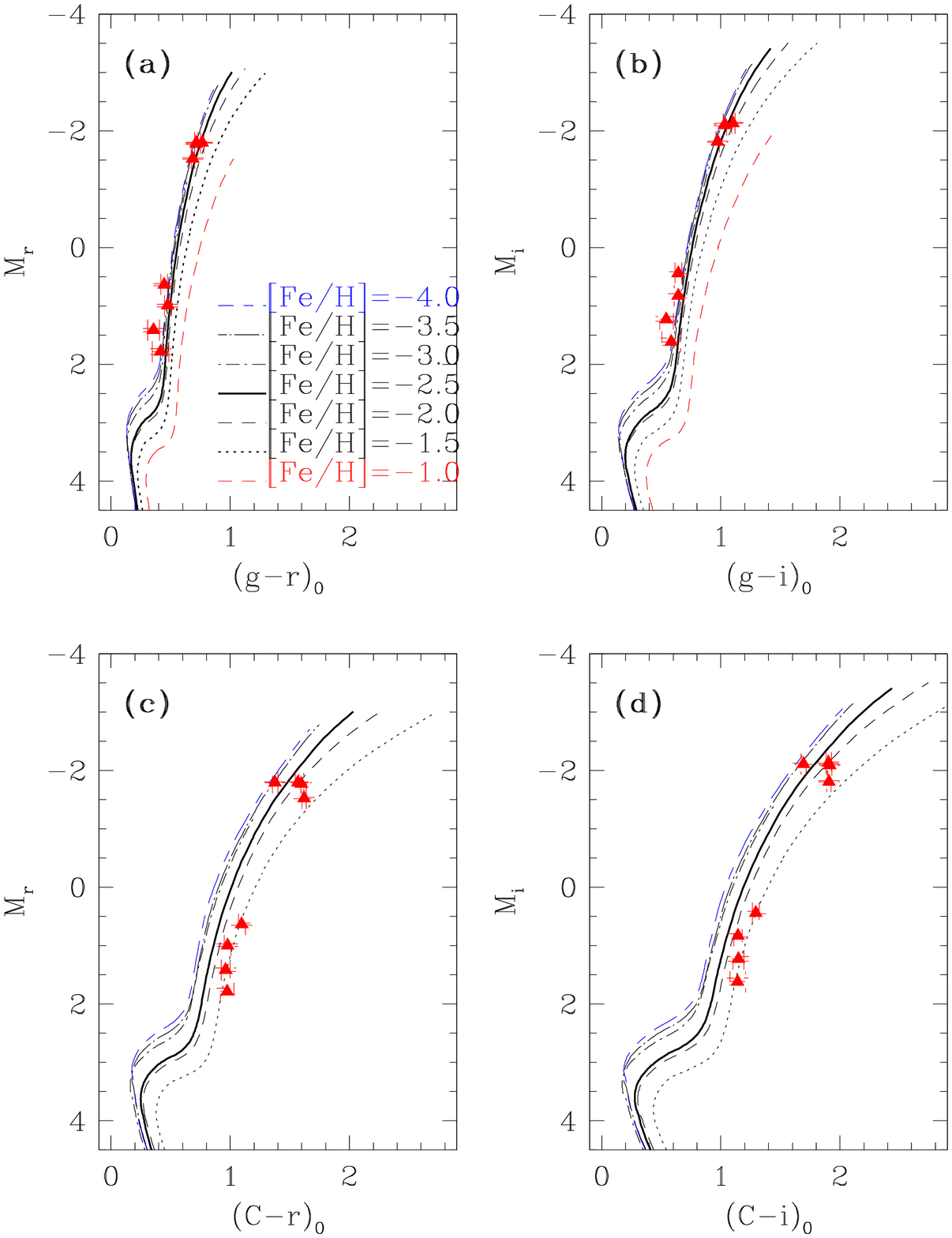}
\vskip0pt
\caption{
All plots show the  Boo I RGB stars as red triangles, with the Dartmouth isochrones for [$\alpha /Fe]=0.2$ and an age of 12~Gyr. Boo I has $E(B-V)=0.02$ and $DM=19.11$.
\bf (a) \rm $M_r \; vs. \; (g-r)$ SDSS CMD.
\bf (b) \rm  $M_i\; vs. \; (g-i)_0$  SDSS CMD.
\bf (c) \rm $M_r \; vs. \; (C-r)$ SDSS/Washington CMD.
\bf (d) \rm  $M_i\; vs. \; (C-i)_0$  SDSS/Washington CMD.}
\label{fig7}
\end{figure}

\begin{figure}
\centering
\includegraphics[scale=0.6]{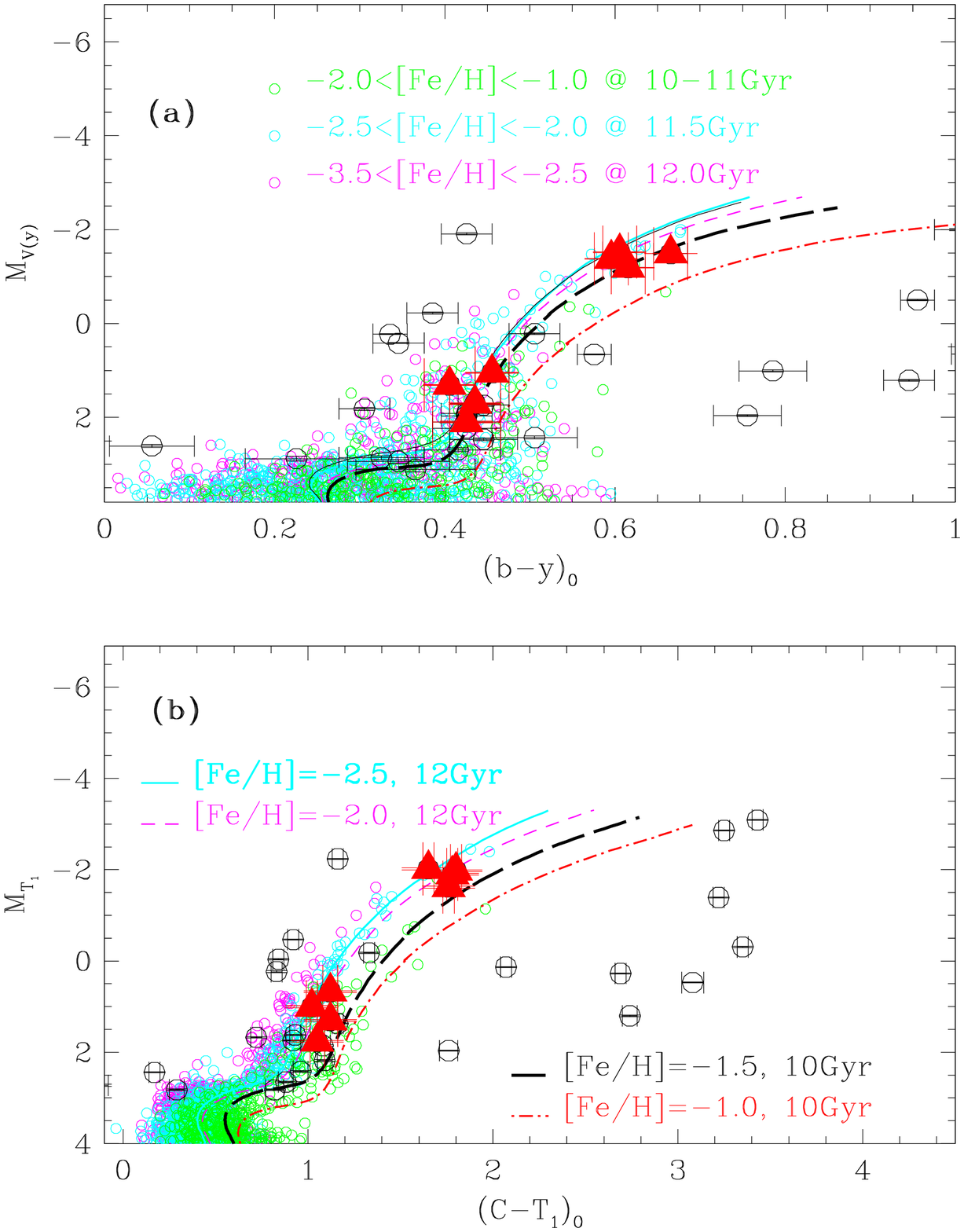}
\vskip0pt
\caption{In both diagrams, the black circles are 34 stars with Str\"{o}mgren and Washington colors, and the red triangles are the 8 RGB stars with SDSS, Str\"{o}mgren and Washington colors.. The magenta points are the oldest, metal-poor stars from the closed-box
model, the cyan points are intermediate-age stars around the average Boo~I [Fe/H] value, and the green points are the youngest, metal-rich(er)
stars, with at least 2\% photometric uncertainties.
\bf (a) \rm $M_{V} \; vs. \; (b-y)$.
\bf (b) \rm $M_{T_1} \; vs. \; (C-T_1)$.
}
\label{fig8}
\end{figure}

\section{Summary and Conclusions}

Taking the most conservative metallicity range of  $-1.5 < [Fe/H] < -2.5$, the Str\"{o}mgren system gives the upper RGB stars  (from HWD)  twice the metallicity resolution of the Washington system at comparable S/N. However, both Str\"{o}mgren- and Washington-color conversions to [Fe/H]-values fail at the lower RGB, due to a combination of temperature and line-blanketing effects., as well as an increase in photometric uncertainties at faint magnitudes. This range in  [Fe/H] produces $\Delta (g-r) \sim 0.1$ dex,  $\Delta (g-i) \sim 0.1$ dex,  $\Delta (g-r) \sim 0.2$ dex, and  $\Delta (C-i) \sim 0.2$ dex, which is still half what we can achieve in the $(C-T_1)$ and the   Str\"{o}mgren system. 

HWB and HWD find that the Washington filters are better suited to dSph population studies than the Sloan filters. The Str\"{o}mgren photometry  is more sensitive to the metallicity than the Washington data for metal-poor systems on the upper RGB. However, the dSphs have so few upper RGB stars, we have to look for a better index than m$_1$ alone. Washington photometry $(C-T_1)$-color is better than both $(V-I)$ and $(B-I)$, both of which are also more affected by reddening.  Combining the m$_1$-index with the $(C-T_1)$ color allows individual stars to have $[Fe/H]_{phot}$  measured to within $\pm 0.25$ dex of spectroscopic values, down the whole RGB. At the MSTO,  we may prefer the $(C-T_1)$ color, to give a longer baseline, but the Str\"{o}mgren isochrones separate further at the MSTO, and are a better age indicator, \it if \rm  we have $\sim$1\% photometry. We recommend that Washington filters are used for systems beyond 100 kpc, or where there no considerable reddening present. We note that the HST WFC3 filter set includes Washington C and the Str\"{o}mgren filters.

\acknowledgements This project used observations obtained with the Apache Point Observatory 3.5-m telescope, which is owned and operated by the Astrophysical Research Consortium. Hughes wishes to thank the APO observing staff for their support and late-night instant-messaging. We also acknowledge financial support from the Kennilworth Fund of the New York Community Trust \& the M. J. Murdock Charitable Trust. We thank Aaron Dotter for running many extra models, Frank Grundahl (for sharing his M92 data), Doug Geisler, Ata Sarajedini, Inese Ivans, Peter Stetson, Jeff Brown, Beth Willman, John Norris, Anna Frebel, Marla Geha, Ricardo Munoz, Kim Venn, and Ryan Leaman for useful discussions. We made use of the SDSS DR7.

\clearpage

\centerline{\Large References}

\begin{flushleft}
    Arnadottir, A.S., Feltzing, S., \& Lundstrom, I  .2010, A\&A, 521, 40

    Belokurov, V.,  Zucker, D. B., Evans, N. W., Wilkinson, M. I., Irwin, M. J., Hodgkin, S., Bramich, D. M., Irwin, J. M., Gilmore, G., Willman, B., Vidrih, S., Newberg, H. J., Wyse, R. F. G., Fellhauer, M., Hewett, P. C., Cole, N., Bell, E. F., Beers, T. C., Rockosi, C. M., Yanny, B., Grebel, E. K., Schneider, D. P., Lupton, R., Barentine, J. C., Brewington, H., Brinkmann, J., Harvanek, M., Kleinman, S. J., Krzesinski, J., Long, D., Nitta, A., Smith, J. A., \& Sneden, S. A. 2006, ApJ, 647, L111

    Bessell, M.S. 2005, A\&A Rev., 43, 293

    Calamida, A., Bono, G., Stetson, P. B., Freyhammer, L. M., Cassisi, S., Grundahl, F., Pietrinferni, A., Hilker, M., Primas, F., Richtler, T., Romaniello, M., Buonanno, R., Caputo, F., Castellani, M., Corsi, C. E., Ferraro, I., Iannicola, G., \& Pulone, L. 2007, ApJ, 670,
400

    Calamida, A., Bono, G., Stetson, P. B., Freyhammer, L. M., Piersimoni, A. M., Buonanno, R., Caputo, F., Cassisi, S., Castellani, M., Corsi, C. E., DallÕOra, M., DeglÕInnocenti, S., Ferraro, I., Grundahl, F., Hilker, M., Iannicola, G., Monelli, M., Nonino, M., Patat, N., Pietrinferni, A., Prada Moroni, P. G., Primas, F., Pulone, L., Richtler, T., Romaniello, M., Storm, J., \& Walker, A. R. 2009, ApJ, 706, 1277

    Carretta, E., \& Gratton, R. G. 1997, A\&AS, 121, 95

    Canterna, R. 1976, AJ, 81, 228 

    Crawford, D.L.,\&  Mander, J. 1966, AJ 71, 114

    di Cecco, A., Becucci, R., Bono, G., Monelli, M., Stetson, P. B., DeglÕInnocenti, S., Prada
Moroni, P. G., Nonino, M., Weiss, A., Buonanno, R., Calamida, A., Caputo, F., Corsi, C. E., Ferraro, I., Iannicola, G., Pulone, L., Romaniello, M., \& Walker, A. R. 2010, PASP, 122, 991 

    Dotter, A., Chaboyer, B., Jevremovic, D., Kostov, V., Baron, E., \& Ferguson, J. W. 2008, ApJS, 178, 89

    Feltzing, S., Eriksson, K., Kleyna, J., \& Wilkinson, M. I. 2009, A\&A, 508, L1

   Geisler, D. 1996, AJ, 111, 480

    Faria, D., Feltzing, S., Lundstrom, I., et al. 2007, A\&A, 465, 357

    Frebel, A., Kirby, E. N., \& Simon, J. D. 2010a, Nature, 464, 72

    Geisler, D., \& Sarajedini, A. 1999, AJ, 117, 308

    Grundahl, F., VandenBerg, D. A., Bell, R. A., Andersen, M. I., \& Stetson, P. B. 2000, AJ, 120, 1884 

    Hilker, M. 2000, A\&A, 355, 994

    Hughes, J., Wallerstein, G., \& Bossi, A. 2008, AJ, 136, 2321

    Hughes, J., Wallerstein, G., \& Dotter, A. 2011, \it in preparation. \rm

   Lai, D., Lee, Y.S., Bolte, M., Lucatello, S., Beers, T.C., Johnson, J.A., Sivaranai, T., \& Rockosi, C.M. 2011, arXiv:astro-ph/1106.2168v1

   Lee, Y.S., et al. 2008a, AJ, 136, 2022 

   Lee, Y.S. et al.  2008b, AJ, 136, 2050

    Martin, N. F., Ibata, R. A., Chapman, S. C., Irwin, M., \& Lewis, G. F. 2007, MNRAS, 380, 281

    Norris, J. E., Gilmore, G., Wyse, R. F. G., Yong, D., \& Frebel, A.
2010a, ApJ, 722, L104 

    Norris, J. E., Wyse, R. F. G., Gilmore, G., Yong, D., Frebel, A.,
Wilkinson, M. I., Belokurov, V., \& Zucker, D. B. 2010b, ApJ,
723, 1632

    Norris, J. E., Wyse, R. F. G., Gilmore, G., Yong, D., Frebel, A., Wilkinson, M. I., Belokurov, V., \& Zucker, D. B. 2010, ApJ, 723, 1632

   Onehag, A., Gustaffsson, B., Eriksson, K., \& Edvardsson, B. 2009, A\&A, 498, 527

   Rutledge, G. A., Hesser, J. E., \& Stetson, P. B. 1997, PASP, 109, 907.

   Siegel, M.H. 2006, ApJ, 649, L83

   Sneden, C., et al., 2003. Astrophys. J., 591, 936 Ð 953

   Str\"{o}mgren, Bengt, 1956,  Vistas in Astronomy, Vol. 2, Issue 1, pp. 1336--1346

   Zinn, R., \& West, M. J. 1984, ApJS, 55, 45

\end{flushleft}

\end{document}